\documentclass[11pt]{article}
\usepackage{latexsym, epsfig, verbatim, amsmath, amssymb, lscape, graphics, threeparttable, color}
\usepackage{mathtools, multirow, rotating}
\usepackage{natbib}

\usepackage{authblk}
\allowdisplaybreaks
\usepackage[margin = 1in]{geometry}
\usepackage{setspace}
\newtheorem{remark}{Remark}

\newcommand{\bd}{{\boldsymbol d}}
\newcommand{\bW}{{\boldsymbol W}}

\newcommand{\wR}{{\widehat R}}
\newcommand{\wV}{{\widehat V}}

\newcommand{\wbeta}{{\widehat\beta}}

\newcommand{\wlambda}{{\widehat\lambda}}

\newcommand{\wrho}{{\widehat\rho}}
\newcommand{\wsigma}{{\widehat\sigma}}
\newcommand{\wOmega}{{\widehat\Omega}}
\newcommand{\tA}{{\widetilde A}}
\newcommand{\tB}{{\widetilde B}}

\begin{document}

%\title{\bf{\large Sample Size Calculation for Active-Arm Trial with Counterfactual Incidence Based on Recency Assay}}
%\author{\normalsize Fei Gao} \affil{\normalsize Fred Hutchinson Cancer Research Center, Seattle, WA, USA.} 
%\author{\normalsize Fan Xia and Kwun Chuen Gary Chan}
%\affil{\normalsize Department of Biostatistics, University of Washington, Seattle, WA, USA.}
%%\affil{Department of Biostatistics, University of Washington, Seattle, WA 98195}
%\date{}
%\bigskip

\title{Sample Size Calculation for Active-Arm Trial with Counterfactual Incidence Based on Recency Assay}
\author[1]{Fei Gao}
\author[2]{David V. Glidden}
\author[3]{James P. Hughes}
\author[1]{Deborah Donnell}
\affil[1]{Vaccine and Infectious Disease Division, Fred Hutchinson Cancer Research Center, Seattle, WA}
\affil[2]{Department of Epidemiology and Biostatistics, University of California, San Francisco, CA}
\affil[3]{Department of Biostatistics, University of Washington, Seattle, WA}
\date{}
\bigskip
\maketitle
\begin{abstract}
The past decade has seen tremendous progress in the development of biomedical agents that are effective as pre-exposure prophylaxis (PrEP) for HIV prevention.
To expand the choice of products and delivery methods, new medications and delivery methods are under development.
Future trials of non-inferiority, given the high efficacy of ARV-based PrEP products as they become current or future standard of care, would require a large number of participants and long follow-up time that may not be feasible.
This motivates the construction of a counterfactual estimate that approximates incidence  for a randomized concurrent control group receiving no PrEP.
We propose an approach that is to enroll a cohort of prospective PrEP users and augment screening for HIV with laboratory markers of duration of HIV infection to indicate recent infections.
We discuss the assumptions under which these data would yield an estimate of the counterfactual HIV incidence and develop sample size and power calculations for comparisons to incidence observed on an investigational PrEP agent.
\end{abstract}

\section{Introduction}

The past decade has seen tremendous progress in the development of biomedical agents that are effective as pre-exposure prophylaxis (PrEP) for HIV prevention \citep{grant2010preexposure}.
To date, use of PrEP by those at risk for HIV infection remains limited \citep{koss2020uptake}.
New medications and delivery methods are under development in the hope that expanding the choice of products and delivery methods will facilitate the scale up of PrEP \citep{baeten2012antiretroviral,thigpen2012antiretroviral,mayer2020discover,landovitz2020hptn083}. 

Trials of investigative PrEP agents should acknowledge the current HIV prevention standard of care.
This is typically done by the use of a randomized active-control non-inferiority design.
However, future trials of non-inferiority, given the high efficacy of ARV-based PrEP products as they become current or future standard of care, would require a much larger number of participants than current trials with longer follow-up times (current trials have sample size 3-5,000 with approximately 2-3 years of follow-up).
Given the epidemiology of the HIV epidemic, enrolling tens of thousands of participants at risk is not feasible \citep{sullivan2018getting, mayer2020barriers}. 

The recognition that we may not be able to conduct fully powered active-control non-inferiority trials for future products motivates a search for alternative study designs that preserve a high evidence standard and are feasible.
One approach could compare HIV incidence among volunteers receiving an investigational PrEP agent to a counterfactual  estimate - one that 
approximates incidence for a randomized concurrent control group 
receiving no PrEP.
A variety of methods for estimating counterfactual incidence based on external data sources have been explored or proposed \citep{glidden2020statistical} including the use of sexually transmitted infection rates \citep{mullick2020correlations}, incidence in placebo arms of recent trial, community HIV surveillance data \citep{mera2019estimation} or use of pharmacology biomarkers \citep{hanscom2019adaptive} when the study includes a tenofovir-based PrEP control group. 

A promising approach is to enroll a cohort of prospective PrEP users and augment screening for HIV with laboratory markers of duration of HIV infection to determine if the infection is recent or not, i.e., an HIV recency test.
We discuss the assumptions under which these data would yield an estimate of the counterfactual HIV incidence and develop sample size and power calculations for comparisons to incidence observed on an investigational PrEP agent.
The basis of our work is an incidence estimator proposed by \cite{kassanjee2012new}.

\section{Approach and Identifiability}

Suppose that we screen $N$ volunteers, not taking PrEP, for a clinical trial of a candidate PrEP agent.
Each person is screened for HIV infection at the screening time (time 0).
Those who are found to be HIV-positive at time 0 are assessed through an HIV recency test to determine whether the HIV infection has a duration of at most $T$ prior to screening.
The determination may not be perfect, as described later in Sections 2.4 and 2.5.
Each person is classified at time 0 as: HIV-negative at 0, HIV-positive in the period $[-T,0]$ (recent infection) or HIV-positive prior to time $-T$.
Suppose at time 0 that $N_+$ subjects are HIV-positive and $N_- \equiv N - N_+$ subjects are HIV-negative.
Suppose that among $N_+$ HIV-positive participants, $N_R$ are found to be recent infections.
Among those HIV-negative participants, suppose $N_{-,enroll}$ are recruited to the active arm of a clinical trial for the candidate PrEP agent and we observe $N_{event}$ incidence cases after $\tau$-year of follow-up.

The efficacy of the candidate PrEP agent is determined by comparing the incidence of subjects receiving the candidate PrEP agent with the underlying incidence of HIV in enrolled participants in the absence of PrEP, i.e., the counterfactual placebo incidence of HIV that we denote as $\lambda_0$.
As described below, we estimate $\lambda_0$ from the recency assay samples.
How closely this will estimate the true counterfactual effect will depend on the alignment of the recency-based incidence estimate and  $\lambda_0$.
It is helpful to contrast some alternative study designs and the estimate of $\lambda_0$ associated with them.

\subsection{Concurrent Randomized Control}

The most rigorous estimate of $\lambda_{0}$ requires randomization of $N_{-,enroll}$ participants between the PrEP intervention and a concurrent
control group and following both groups for HIV incidence over $[0,\tau]$ (shown in Figure 1).
This would yield a non-PrEP incidence ($\lambda_0$) in a population with subjects who are (i) HIV-negative, (ii) eligible for and (iii) consenting to the PrEP intervention at time 0.
The concurrent randomized control trial provides an unbiased estimate of the treatment efficacy under minimal assumptions.
We can judge other non-PrEP estimates by how well they replicate the randomized scenario.

\begin{figure}[h]
\centering
\includegraphics[width=3.5in]{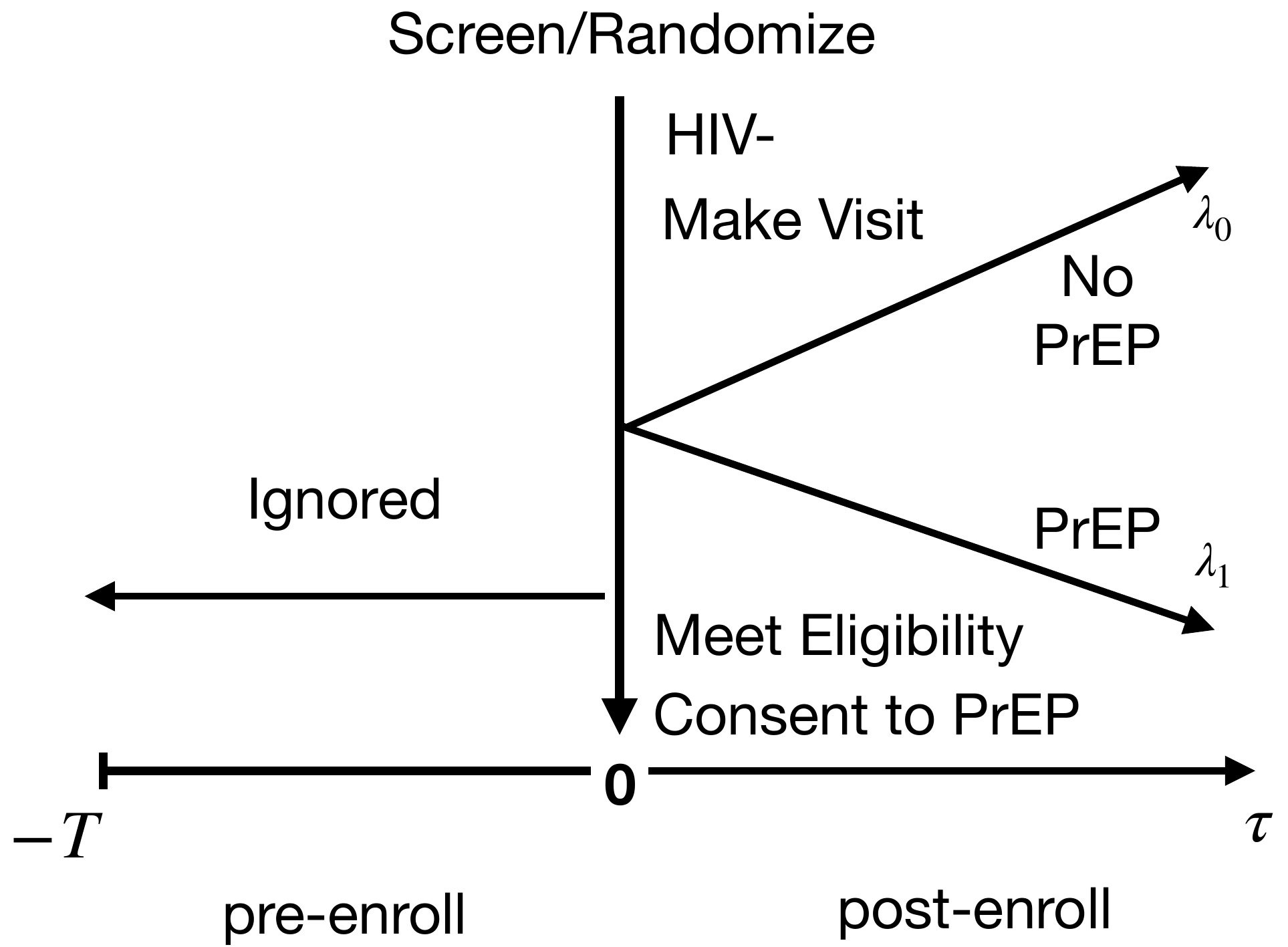}
\caption{Concurrent randomized control trial.}
\label{fig:simu_lambda_true}
\end{figure}

\subsection{Cross-over Trial}

An alternative approach is to assess a PrEP agent in a cross-over trial (shown in Figure 2).
Suppose that we enroll at-risk HIV-negative participants at time $-T$ who received no PrEP and ask them to return at time 0.
They are again assessed at time 0 and those who satisfies (i), (ii), and (iii) are enrolled to receive the PrEP intervention.
This on-study incidence could be compared to the HIV incidence over $[-T,0]$ among those who returned at time 0 and meet (ii) and (iii).
% We assumed no HIV testing between $-T$ and 0, so pre-PrEP incidence is interval-censored.

\begin{figure}[h]
\centering
\includegraphics[width=3.5in]{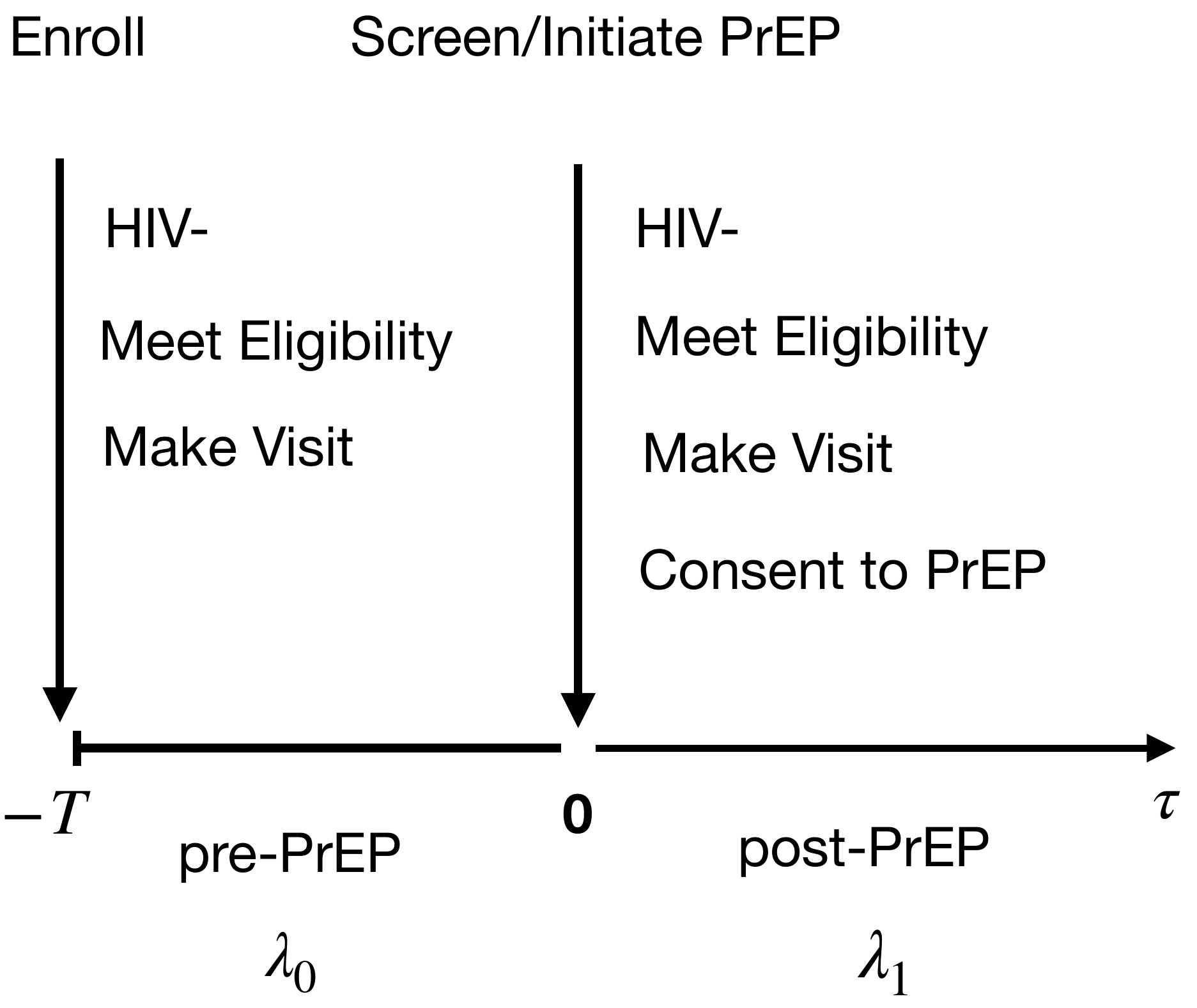}
\caption{Cross-over trial.}
\label{fig:simu_lambda_true}
\end{figure}

The cross-over trial has the potential for certain biases relative to the concurrent randomized control trial.
Since the cross-over trial constructs a closed cohort consisting of subjects who are HIV-negative and at risk at $-T$, there is a selection effect if HIV risk is highly variable in the population: those enrolled to receive PrEP intervention have an average lower risk.
There may also be structural time trends in HIV risk (e.g., treatment as prevention) which will render the two periods non-comparable.
People may test positive before time $0$ and may not present for screening at time $0$ because they know their status.
Finally, people may initiate PrEP in $[-T,0)$.

In practice, it may also be difficult to assess (ii) and (iii) for those who were HIV-positive at time 0.
A sufficient condition for the pre-PrEP incidence to coincide with the concurrent control estimand is that among those HIV negative subjects at $-T$ there is no systematic difference between HIV risks of
those infected and uninfected at time $0$, incidence is constant over $[-T,\tau)$ in those not taking PrEP, there is no HIV testing or PrEP use between $-T$ and 0, the attendance to screening at time 0 is independent of HIV status, and the criterion (ii) and (iii) can be honestly assessed for subjects who return at time 0.

\subsection{Perfect Recency Test at Enrollment}
Suppose that we screen a group of individuals, neutral to HIV status, who would be eligible for and willing to initiate the PrEP intervention.
These individuals are tested at time 0 for HIV infection and for HIV recency over the period $T$.
Figure 3 illustrates such a trial.
Assume there is no misclassification in recency assessment such that we accurately determine whether an HIV-positive subject was infected during $[-T,0)$.
Let $N_R$ be number of recent infections by this perfect recency test.

\begin{figure}[h]
\centering
\includegraphics[width=3.5in]{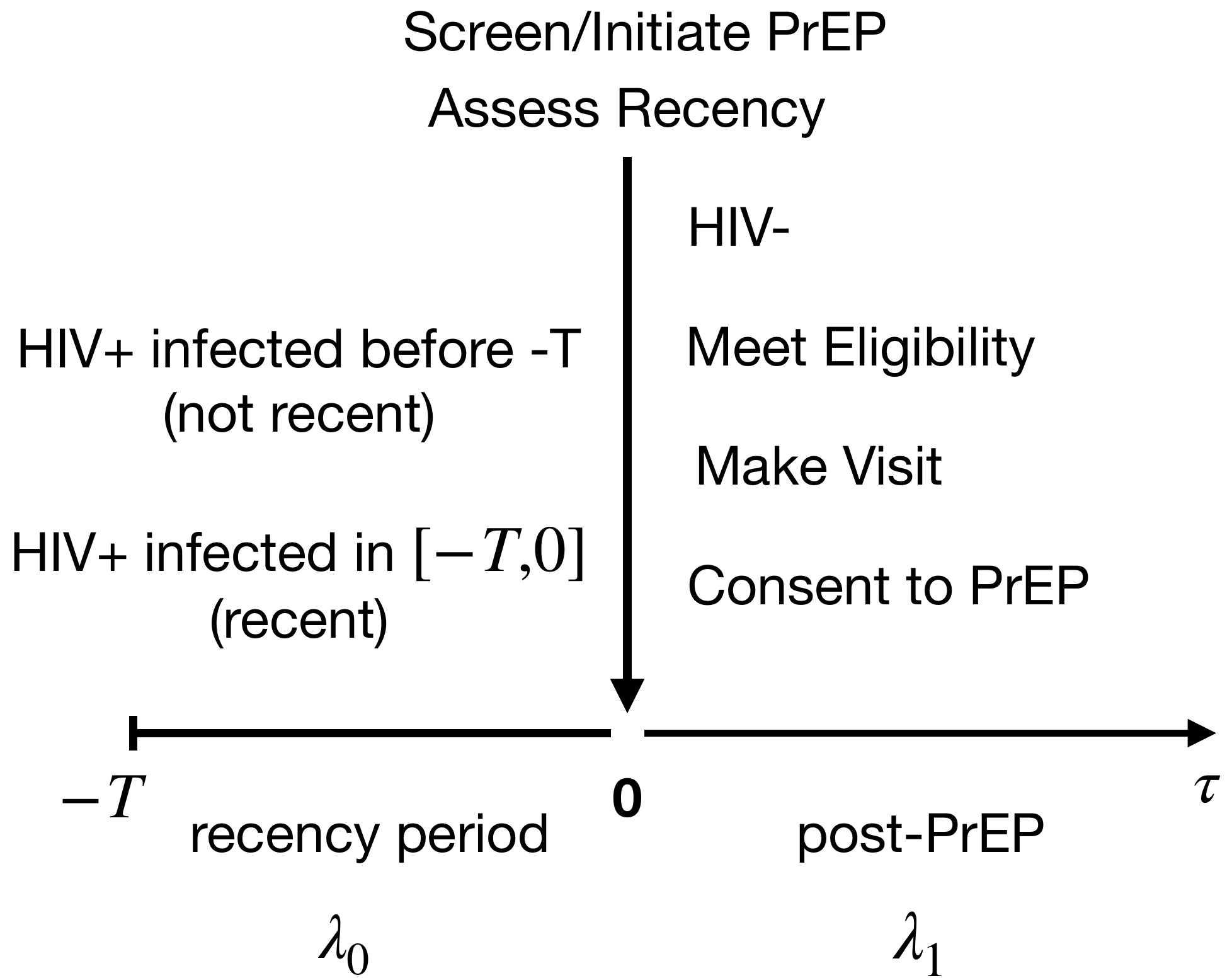}
\caption{Trial with perfect recency test at enrollment.}
\label{fig:simu_lambda_true}
\end{figure}

This design mimics the cross-over trial and the difference is that the data are collected retrospectively rather than prospectively.
Indeed, the recency test approach is based on subjects from an open cohort, with evolving at-risk population, instead of a closed cohort as in the cross-over trial.
This will alleviate the selection bias induced by the cross-over design, since the high-risk subjects in the population will be ``refreshed''.
It will also alleviates some of the concerns about honest assessment of (ii) and (iii) associated with the cross-over design.
%The estimates will coincide with those from the cross-over trial if all subjects assessed at time 0 are indeed HIV-negative, at risk, and enrolled to the cross-over trial at time $-T$.

The (perfect) recency test approach  yields multinomial data with counts and probabilities shown below.
\begin{center}
\begin{tabular}{lccc}
   Category  & Count & Probability & Approximate Probability \\
   HIV+ (Not Recent) & $N_+ - N_R$ & $ p - (1-\pi)F_0(T)$ & $p - (1-p)\lambda_0T$ \\
   HIV+ (Recent) & $N_R$ & $(1-\pi)F_{0}(T)$ & $(1-p)\lambda_0T$\\
    HIV-  & $N_-$ & $ 1-p$ & $ 1-p$ \\
\end{tabular}
\end{center}
Here, $p$ and $\pi$ are the prevalence of HIV+ at time $0$ and $-T$, respectively, and $F(\cdot)$ is
the cumulative distribution function of time to HIV infection.
We assume constant incidence over $[-T,\tau]$ and this incidence is $\lambda_{0}$.
Thus, $F_{0}(t) = 1-\exp(-t\lambda_0)\approx t\lambda_0$ for $t\le T$ when $\lambda_0$ is small and $T$ is relatively short.
We assume constant prevalence over $[-T,0]$ such that $\pi = p$.
Estimation of $\lambda_{0}$ by maximizing the likelihood based on approximate probabilities is particularly simple in this case and is given by $ N_R / (N_-T)$.

\subsection{Recency Test with False Negative Recency}

In practice, all recency tests allow for false negative recency, i.e., participants that are infected less than $T$ years identified as long-infected.
Let $\phi(t)$ be the probability that an individual infected $t$ years ago is identified as recently infected by the test, i.e., true positive rate.
We suppose for now that $\phi(t)$ vanishes for $t>T$, i.e, there is no false positive probability.

Suppose that incidence is constant over $[-T,\tau]$ and prevalence is constant over $[-T,0]$.
When $\lambda_0$ is small and $T$ is relatively short, the observed multinomial data with counts and probabilities are shown below.
\begin{center}
\begin{tabular}{lcc}
   Category  & Count & Approximate Probability  \\
   HIV+ (Not Recent) & $N_+ - N_R$ & $p - (1-p)\lambda_0\Omega_T$ \\
   HIV+ (Recent) & $N_R$ & $(1-p)\lambda_0\Omega_T$ \\
    HIV-  & $N_-$ & $ 1-p$ \\
\end{tabular}
\end{center}
Here, $\Omega_T = \int_0^T\phi(t)dt$ is the mean window period of the recency test.
The maximum likelihood estimator for $\lambda_0$ (based on approximate probabilities) is given by $N_R / (\Omega_T N_-)$, which is the snapshot estimator in \cite{kaplan1999snapshot}.
Note that compared to the estimator in Section 2.3, this estimator has an additional element $\Omega_T/T$ in the denominator, reflecting average false negative rate.

\subsection{Counterfactual Placebo HIV Incidence Estimator}
In practice, assays are imperfect and we must allow for both false 
positive and false negative recency classifications among HIV+ persons, that is $\phi(t)> 0$ for $t>T$.
Suppose that $\phi(t)$ is approximately constant for $t>T$ and we let $\beta_T$ be the constant value.
In that case, $\Omega_T$ is called mean duration of recent infection (MDRI) and $\beta_T$ coincides with false-recent rate (FRR) defined in \cite{kassanjee2012new}, which is the probability that a randomly chosen person infected for more than $T$ years is test-recent.
Then, the observed multinomial data with counts and probabilities are shown below.
\begin{center}
\begin{tabular}{lcl}
   Category  & Count & Approximate Probability  \\
   HIV+ (Not Recent) & $N_+ - N_R$ & $p - (1-p)\lambda_0\Omega_T-\{p - (1-p)\lambda_0T\}\beta_T$ \\
   HIV+ (Recent) & $N_R$ & $(1-p)\lambda_0\Omega_T +\{p - (1-p)\lambda_0T\}\beta_T$ \\
    HIV-  & $N_-$ & $ (1-p)$ \\
\end{tabular}
\end{center}
The maximum likelihood estimator for $\lambda_0$ (based on approximate probabilities) is given by
\begin{equation}
\wlambda_0 = \frac{N_R - \beta_T N_+}{N_-(\Omega_T - \beta_TT)},\label{equ:lambda0_est}
\end{equation}
which is the HIV incidence estimator proposed in \cite{kassanjee2012new}.
Compared to the estimator in Section 2.3, this estimator has a different numerator reflecting false recent adjustment and an additional element $\Omega_T/T - \beta_T$, reflective adjustment on false positive and false negative rates.
In practice, we replace $\Omega_T$ and $\beta_T$ in (\ref{equ:lambda0_est}) by $\wOmega_T$ and $\wbeta_T$, which are the estimated MDRI and FRR, respectively. 

For the remainder of this paper, we will adopt the \cite{kassanjee2012new} estimator.
This estimator \eqref{equ:lambda0_est} (with $\wOmega_T$ and $\wbeta_T$) will be consistent as $n \to \infty$, subject to consistency of $\wbeta_T$ and $\wOmega_T$ and similar conditions as required in Sections 2.2-2.4.
Particularly, it would require (a) no systematic difference between HIV risks of those infected recently and those eligible for the trial, (b) incidence is constant over $[-T,\tau)$ in those not taking PrEP, (c) willingness to HIV screening is independent of HIV status, (d) prevalence is constant over $[-T,0]$, and (e) the criterion (ii) and (iii) can be honestly assessed for screened subjects.

\subsection{Treatment Efficacy Evaluation}
The HIV incidence in the active-arm trial, $\lambda_1$, can be estimated by 
\begin{equation}
\wlambda_1 = \frac{N_{event}}{\tau N_{-,enroll}}.\label{equ:lambda1_est}
\end{equation} 
Write $R = \lambda_1/\lambda_0$ as the incidence ratio.
Then, the efficacy of the active treatment, represented by the percentage incidence reduction $\rho \equiv 1-R$, can be estimated by $\wrho \equiv 1-\wR$, where $\wR = \wlambda_1 / \wlambda_0$.

This paper will focus on developing a variance expression for $\log\wR$ based on (\ref{equ:lambda0_est}) and (\ref{equ:lambda1_est}).
It will be used to develop power calculations for testing hypothesis based on $\wrho$.

\section{Asymptotics and Power}

\subsection{Inference for Incidence Estimators and Treatment Efficacy}
Let $P_R = \beta_T + \frac{\lambda_0(1-p)}{p}(\Omega_T - \beta_TT)$, where $p$ is the HIV prevalence at time 0.
$P_R$ is the probability of test-recent among HIV-positive subjects.
The number of test-recent subjects $N_R$ can be viewed as generated from a Binomial distribution with size $N_+$ and success probability $P_R$, while the number of HIV-positive subjects $N_+$ is from a Binomial distribution with size $N$ (total screened subjects) and success probability $p$.
Applying the delta method (with details shown in Appendix \ref{append:deriv_var}), we obtain that the log estimated incidence has asymptotic distribution
\[\log\wlambda_0\sim N\left(\log\lambda_0, V_0(\lambda_0)\right),\]
where $V_0(\lambda_0) = \gamma_{00}(\lambda_0)/N+\gamma_{01}(\lambda_0)$,
\begin{align*}
\gamma_{00}(\lambda_0) =& \frac{1}{p}\left\{\frac{P_R\left(1-P_R\right)}{\left(P_R-\beta_T\right)^2} +\frac{1}{(1-p)} +\frac{(1-p)\sigma_{\wbeta_T}^2 }{\left(P_R-\beta_T\right)^2}\right\},\\
\gamma_{01}(\lambda_0) =& \frac{\sigma_{\wOmega_T}^2}{\left(\Omega_T - \beta_TT\right)^2} +\sigma_{\wbeta_T}^2\left\{\frac{\left(\Omega_T - P_RT \right)^2}{\left(P_R-\beta_T\right)^2\left(\Omega_T - \beta_TT\right)^2}\right\},
\end{align*}
and $\sigma_{\wOmega_T}^2$ and $\sigma_{\wbeta_T}^2$ are the variances of $\wOmega_T$ and $\wbeta_T$, respectively.
This formula is slightly different from the formulas in \cite{kassanjee2012new} and in R package `inctools' , where the last term of $\gamma_{00}(\lambda_0)$ is neglected.
However, the difference is minimal when $\sigma_{\wbeta_T}^2$ is small.
To estimate $V_0(\lambda_0)$, we replace the expected values by their estimators to obtain the variance estimator 
\begin{align*}
\wV_0 =&\frac{N_R\left(N_+-N_R\right)}{N_+\left(N_R-N_+\wbeta_T\right)^2} +\frac{N}{N_+N_-} + \wsigma_{\wbeta_T}^2\frac{N_+(N-N_+) }{N\left(N_R-N_+\wbeta_T\right)^2} \nonumber\\
& + \frac{\wsigma_{\wOmega_T}^2}{\left(\wOmega_T - \wbeta_TT\right)^2}+\wsigma_{\wbeta_T}^2\left\{\frac{N_+\wOmega_T - N_RT}{\left(N_R-N_+\wbeta_T\right)\left(\wOmega_T - \wbeta_TT\right)}\right\}^2.
\end{align*}

The number of incidence cases in the active-arm trial $N_{event}$ follows a Poisson distribution with mean $\lambda_1 N_{-,enroll}\tau$.
Then, the log estimated incidence has asymptotic distribution
\[\log\wlambda_1 \sim N\left( \log\lambda_1, V_1(\lambda_1)\right),\]
where $V_1(\lambda_1) = \gamma_1(\lambda_1)/N$,
\[\gamma_1(\lambda_1) = \frac{1}{\lambda_1(1-p)r\tau},\]
and $r$ is the probability of enrollment among HIV-negative subjects at time 0.
We replace the expected values by their estimators to obtain an estimator for $V_1(\lambda_1)$
\[\wV_1 = \frac{1}{N_{event}}.\]

In Appendix \ref{append:deriv_var}, we find the asymptotic variance of $\log\wR$ is equal to $\wV_0 + \wV_1$.
Then, a 95\% confidence interval for $\rho$ can be constructed as
\[\left(1 - \frac{\wlambda_1}{\wlambda_0}\exp\left(z_{0.975}\sqrt{\wV_0+\wV_1}\right),1 - \frac{\wlambda_1}{\wlambda_0}\exp\left(-z_{0.975}\sqrt{\wV_0+\wV_1}\right)\right),\]
where $z_c$ is the $c$-quantile of the standard normal distribution.

\subsection{Sample Size Determination} 
We would like to determine the sample size for testing $H_0: R = R_0$, with significance level $\alpha$ and power $\beta$ against a specific alternative $H_1: R = R_1$.
This is equivalent to testing $H_0: \log R = \log R_0$ versus the specific alternative $H_1: \log R = \log R_1$.
Based on the asymptotic distribution of $\log\wR$, we consider the $Z$-statistic
\begin{equation}
Z = \frac{\log\wR-\log R_0}{\sqrt{\wV_0 + \wV_1}}.\label{equ:Z_stat}
\end{equation}
Under the null hypothesis, the $Z$ statistic is asymptotically $N(0,1)$ distributed.
Under the alternative hypothesis, the mean of the $Z$ statistic is
\begin{align*}
E(Z)\approx& \frac{\log R_1-\log R_0}{\sqrt{V_0(\lambda_0) + V_1(\lambda_0R_1)}},
\end{align*}
but the asymptotic variance of $Z$ departs from 1 significantly.
In Appendix \ref{append:deriv_Zstat}, we show the derivation of the asymptotic variance of $Z$ under $H_1: R = R_1$, denoted as $V_{R_1}$.
To attain $\alpha$-significance level, the cut-off for rejecting null hypothesis is set to $z_{1-\alpha/2}$.
To attain power $\beta$, we need
\[\Pr\left(W_{R_1} +\frac{\log R_1-\log R_0}{\sqrt{\{\gamma_{00}(\lambda_0) + \gamma_1(\lambda_0R_1)\}/N + \gamma_{01}(\lambda_0)}}>z_{1-\alpha/2}\right)\ge \beta,\]
where $W_{R_1}$ is an independent normal random variable with mean 0 and variance $V_{R_1}$ (expression given in Appendix B),
such that the sample size is given by 
\begin{equation}
N = \frac{\gamma_{00}(\lambda_0)+\gamma_1(\lambda_0R_1)}{\left\{\dfrac{\log R_1-\log R_0}{z_{1-\alpha/2} + \sqrt{V_{R_1}}z_\beta}\right\}^2 - \gamma_{01}(\lambda_0)}.\label{equ:n_log}
\end{equation}
\begin{remark}
Note that the variance of $\log\wR$, $V_0(\lambda_0) + V_1(\lambda_0R_1)$, will not go to zero as the total screening sample size $N$ goes to infinity.
Particularly, as $N$ goes to infinity, the asymptotic variance converges to $\gamma_{01}(\lambda_0)$, which is a weighted sum of the variabilities from $\wOmega_T$ and $\wbeta_T$.
Therefore, we are not able to to achieve $\beta$-power for alternative hypothesis ($H_1:R=R_1$) with 
\[\log R_1>\log R_0 + \sqrt{\gamma_{01}(\lambda_0)}\left(z_{1-\alpha/2} + \sqrt{V_{R_1}}z_\beta\right).\]
\end{remark}

\section{Numerical Results}
\subsection{Sample Size Calculation}
We first consider the setting of a hypothetical trial for men who have sex with men (MSM) and transgender women (TGW), with trial participants recruited from different regions.
Specifically, we mimic the composition of the screening population of the HIV Prevention Network (HPTN) 083 study \citep{landovitz2020hptn083}.
Details on the incidence, prevalence are shown in Table \ref{tab:setting}, together with the MDRI, MDRI relative standard error (RSE), and FRR of the recency test based on LAg Avidity (Sedia HIV-1 LAg Avidity EIA; Sedia Biosciences Corporation, Portland, OR, USA) OD$_n$ $\le$ 1.5 and viral load $>$ 1000 copies/ml with cutoff $T = 2$ years \citep{grebe2019post}, where an Estimated dates of detectable infection (EDDIs) offset of 16 days was applied to the MDRI for using 4th generation assay for HIV diagnosis \citep{facente2020estimated}.

\begin{table}[htbp]
\protect\caption{Composition of subtypes for hypothetical MSM trial.}\label{tab:setting}
\centering
\begin{tabular}{lcccccccc}
\hline\hline
Region	&	Proportion	&	Incidence	&	Prevalence	&	Subtype	&	MDRI (days)	&	MDRI RSE	&	FRR	\\
US-Black	&	18.5	\%&	5.9	\%&	15	\%&	B	&	\multirow{5}{*}{142}	&	\multirow{5}{*}{10}	&	\multirow{5}{*}{1.5\%}	\\
US-Other	&	18.7	\%&	1.3	\%&	15	\%&	B	&		&		&		\\
Brazil	&	17.5	\%&	5	\%&	15	\%&	B	&		&		&		\\
Peru	&	18.2	\%&	3.5	\%&	15	\%&	B	&		&		&		\\
Buenos Aires	&	7.3	\%&	6.4	\%&	15	\%&	B	&		&		&		\\\\
Cape Town	&	3.3	\%&	4.7	\%&	25	\%&	C	&	118	&	7	&	1.0\%	\\\\
Bangkok	&	9.1	\%&	5.2	\%&	15	\%&	A/E	&	\multirow{3}{*}{NA}	&	\multirow{3}{*}{NA}	&	\multirow{3}{*}{NA}	\\
Chiang Mai	&	3.1	\%&	8.2	\%&	15	\%&	A/E	&		&		&		\\
Hanoi	&	4.4	\%&	4	\%&	15	\%&	A/E	&		&		&		\\
\hline
\end{tabular}
\protect
\end{table}

Based on this combination of subtypes, the overall incidence and prevalence of HIV are 4.4\% and 15\%, respectively.
Since the property of the recency test for subtype A/E is not available, we approximately calculate the overall performance of the recency test by weighting the MDRI and FRR among subtypes B and C, to obtain an MDRI of 141 days with RSE 10\% and an FRR of 1.5\%.
We assume 25\% RSE for FRR, all HIV-positive subjects will receipt recency test, and 85\% of HIV-negative subjects will be enrolled to received active treatment.
We consider $\tau = 1$ or 2 to examine the effect of follow-up time on sample size.

We consider the null hypothesis $H_0: R = 0.5$, i.e., the active treatment is 50\% effective for preventing HIV infection and the alternative hypothesis $H_1: R = 0.15$, i.e., the active treatment prevents 85\% of HIV infections.
The null hypothesis reflects the fact that future products should be highly effective such that effectiveness compared to placebo would be expected to exceed 50\%.
The required total screening sample sizes for $\alpha = 0.05$ and $\beta = 0.9$ are given in Table \ref{tab:samplesize}.
We also display the expected number of events under the alternative hypothesis.
If all subjects are followed for one year in the active-arm trial, then 2000 subjects are needed for screening, leading to 30.9 expected recency-test-positive subjects and 9.4 expected incidence cases in the trial.
If the follow-up time is extended to two years, then about 450 fewer subjects are needed for screening, with fewer expected recency-test-positive subjects but more expected incidence cases observed in the trial.
\begin{table}[htbp]
\protect\caption{Total screening sample sizes and expected number of events for MSM and TGW population under alternative hypothesis.}\label{tab:samplesize}
\centering
\begin{tabular}{ccccccc}
\hline\hline
Follow-up Year	&	Screening	&	\multicolumn{2}{c}{Recency Test}			&&	\multicolumn{2}{c}{Active-Arm Trial}			\\\cline{3-4}\cline{6-7}
$\tau$	&	$N$ 	&	$E(N_+)$	&	$E (N_R)$	&&	$E(N_{-,enroll})$	&	$E(N_{event})$	\\\hline
1	&	2000	&	306.7	&	30.9	&&	1439.3	&	9.4	\\
2	&	1545	&	236.9	&	23.9	&&	1111.9	&	14.6	\\
\hline
\end{tabular}
\protect
\end{table}

We also consider another setting for a population of young women in sub-Saharan Africa, mimicking the population enrolled in HPTN 084.
The population is dominated by subtype C, such that the mean MDRI of 119 days with 7\% RSE and FRR of 1.0\% are used.
The overall incidence and prevalence of HIV are 3.5\% and 25\%, respectively.
We consider the same hypothesis testing $H_0: R = 0.5$ vs. $H_1: R = 0.15$.
The required total screening sample sizes for $\alpha = 0.05$ and $\beta = 0.9$ are given in Table \ref{tab:samplesize2}.
The total screening sample sizes are larger, mostly driven by lower HIV incidence in the sub-Saharan Africa women population.

\begin{table}[htbp]
\protect\caption{Total screening sample sizes and expected number of events for sub-Saharan Africa women population under alternative hypothesis.}\label{tab:samplesize2}
\centering
\begin{tabular}{ccccccc}
\hline\hline
Follow-up Year	&	Screening	&	\multicolumn{2}{c}{Recency Test}			&&	\multicolumn{2}{c}{Active-Arm Trial}			\\\cline{3-4}\cline{6-7}
$\tau$	&	$N$ 	&	$E(N_+)$	&	$E (N_R)$	&&	$E(N_{-,enroll})$	&	$E(N_{event})$	\\\hline
1	&	3811	&	952.8	&	43.6	&&	2429.5	&	12.8	\\
2	&	3236	&	809.0	&	37.0	&&	2063.0	&	21.7	\\
\hline
\end{tabular}
\protect
\end{table}

\subsection{Simulation Studies}
The proposed sample size calculation procedure is based on asymptotic theory of the estimators.
To evaluate if the proposed testing procedure has desired type-I error and power in finite samples, we conduct simulation studies.
For a given total screened sample size $N$ and given risk ratio $R$, we use the following simulation procedure.
\begin{enumerate}
\item We generate $N_+$ from $Bin(N, p)$, where $p$ is prevalence and calculate $N_- = N- N_+$.
\item We generate $N_R$ from $Bin(N_+,P_R)$, where $P_R = \beta_T + \frac{\lambda_0(1-p)}{p}(\Omega_T - \beta_TT)$.
We generate $\wbeta_T\sim N(\beta_T,\sigma_{\wbeta_T}^2)$, and $\wOmega_T\sim N(\Omega_T,\sigma_{\wOmega_T}^2)$.
\item We generate $N_{-, enroll}$ from $Bin (N_-,r)$, where $r$ is the proportion of HIV-negative subjects enrolled to the trial.
We generate $N_{event}$ from Poisson distribution with mean $\tau\lambda_1N_{-,enroll}$, where $\tau$ is the follow-up year and $\lambda_1=\lambda_0 R$.
\item We calculate the incidence estimates $\wlambda_0$ and $\wlambda_1$ by (\ref{equ:lambda0_est}) and (\ref{equ:lambda1_est}), respectively, and calculate their variance estimators $\wV_0$ and $\wV_1$.
\item We calculate $Z$ by formula (\ref{equ:Z_stat}).
\end{enumerate}
For each setting, we simulate the data using the calculated sample size as in Table \ref{tab:samplesize} under the null hypothesis $R=R_0=0.5$ or alternative hypothesis $R=R_1 = 0.15$.
We evaluate the type-I error and power by calculating the average rejection ($|Z| >z_{0.975}$) probabilities under null and alternative hypotheses, respectively.
Table \ref{tab:simu_power} shows the simulation results based 10,000 replicates.
The proposed testing procedure has empirical type-1 error smaller than 0.05 and empirical power close to 0.9 (nominal level).

\begin{table}[htbp]
\protect\caption{Simulation Results for Type-I Error and Power}\label{tab:simu_power}
\centering
\begin{tabular}{ccccccccccc}
\hline\hline
\multirow{2}{*}{Setting}	&	Follow-up Year	&	Screening	&	\multirow{2}{*}{Type-I Error}	&	\multirow{2}{*}{Power}	\\
	&	$\tau$	&	$N$ 	&		&		\\\hline
\multirow{2}{*}{MSM and TGW}	&	1	&	2000	&	0.044	&	0.882	\\
	&	2	&	1545	&	0.042	&	0.889	\\\\
\multirow{2}{*}{Women}	&	1	&	3811	&	0.035	&	0.859	\\
	&	2	&	3236	&	0.038	&	0.869	\\
\hline
\end{tabular}
\protect
\end{table}

\section{Discussion}

In this paper, we derived sufficient conditions for estimating counterfactual incidence in active-arm trial based on recency assay and proposed methods for calculating sample sizes.
Our results suggest that future one-arm trials with counterfactual placebo incidence based on a recency assay can be conducted with reasonable total screening sample sizes and adequate power to determine treatment efficacy.

Counterfactual incidence based on a recency assay is closely related to a cross-over design in which a cohort has a pre-PrEP period used as a comparator for a post-PrEP period.
%The major advantage of the proposed approach, like the cross-over design, is that the population which enters the PrEP trial overlaps with population in which counterfactual placebo incidence is estimated.
The use of recency designs allows a cohort to be constructed retrospectively among those who screen for a trial, such that the assumption on similar HIV risks between subjects that contribute to two estimators is more plausible due to the open-cohort nature of the recency approach.

An important context for this alternative design approach is the expectation that future products will be required to be highly effective relative to placebo, given high efficacy of  previously tested ARV-based PrEP products (FTC/TDF, F/TAF \citep{mayer2020emtricitabine} and CAB-LA \citep{landovitz2020hptn083}); effectiveness compared to placebo would be expected to exceed 50\%.
Additionally, because of the existence of highly effective PrEP, it is also assumed that effectiveness below some threshold, e.g., 50\% would be considered unacceptable.
This context – essentially the same that mandates the use of randomized active-control non-inferiority designs – also supports the assumption that substantial differences in HIV incidence are expected between no-PrEP and PrEP groups.
Large observed differences in HIV incidence, as used in our examples, are also likely to be robust to small deviations from our assumptions.

The proposed approach requires a population which is not currently engaged in HIV care or active prevention and this would be a major shift in how populations are screened for HIV prevention studies.
A major objective of the trial will not only be reaching a target number of on-study infections but also reaching a minimum number of recent infections.
Some aspects of the screening phases of the trial would have to be adapted to accomplish the dual aims of estimating on-PrEP and off-PrEP HIV incidence.

Recency assays will misclassify individuals and good estimates of these error rates are a key part of this estimation.
Further, recency assay misclassification rates are a source of variability that are not reduced by the size of the trial.
Improved assay performance or less uncertainty about the misclassification rates will improve incidence estimation.
The choice of $T$ from the assay requires importance considerations.
If the period $T$ is chosen to be short, there will be few recent infections and power is reduced.
If the period $T$ is chosen to be long, the assumptions of the approach can become more implausible and effects of violations can be magnified. 

An estimate of relative efficacy not based on a randomized comparison necessarily makes assumptions about HIV risk in the participants contributing to the two estimates.
In our case our strongest assumption is that the distribution of HIV risk exposure remains the same in the group (and period) assessed for cross-sectional incidence and the group (and period) receiving active product.
We also assume willingness to enter trial screening is independent of HIV status.
Both of these assumptions relate to the characteristics of participants at the time of trial entry, and point to the need for careful attention to screening processes and eligibility criteria. 

We calculate the required screening sample size by a closed form formula based on normal approximations of log incidence estimate distributions, making use of the unexpected fact that $\wlambda_0$ and $\wlambda_1$, the incidence estimates from the recency assay and the active-arm trial, are asymptotically independent.
Alternatively, we can use the exact distributions of these variables and determine the required sample size by simulations.
Noted that complete exploration of the design space via simulations is intensive, both computationally and in terms of programming.
The closed form formula we propose has nice properties in realistic settings such that the empirical power is close to the desired power.

In estimating counterfactual incidence based on recency assay, we made use of estimator proposed by \cite{kassanjee2012new}, which maximizes the likelihood based on approximate probabilities.
A natural extension is to explore the likelihood function without approximations or constant incidence and prevalence requirements.
Based on such a likelihood, we may further consider Bayes-based approaches, where external information (e.g., incidence or prevalence estimated from other sources) can be incorporated.

Our formulation of objectives has focused on one-arm trial design with counterfactual placebo based on recency assay, as a first step to assess necessary assumptions and provide power analysis.
Randomized active-control non-inferiority trial may also be suggested to provide valuable safety comparison and assessment of comparative effectiveness.
Our next step is to explore statistical and practical issues in combining an active-control non-inferiority trial combined with a recency assay counterfactual incidence estimate.

\bibliography{cross}

\newpage
\appendix

\section{Derivation of Asymptotic Variances}\label{append:deriv_var}
Note that $N$, $N_+$, and $N_-$ are the numbers of total screened, HIV-positive, and HIV-negative subjects, $p$ is HIV prevalence, $r$ is the proportion of HIV-negative subjects enrolled to the trial, $N_{-,enroll}$ is the number of HIV-negative subjects enrolled to the trial, and $\tau$ is the follow-up time, and $\wbeta_T$ and $\wOmega_T$ are the estimated false recency rate and MDRI for the recency assay.

Write $\bW = (N_R - N_+\wbeta_T,N_+, \wOmega_T -\wbeta_TT, N_{event},N_{-,enorll})^{\rm T}$.
Then, the estimators $\wlambda_0$ in (\ref{equ:lambda0_est}) and $\wlambda_1$ in (\ref{equ:lambda1_est}) can be written as 
\[\wlambda_0 = \frac{W_1}{(N-W_2)W_3}\]
and 
\[\wlambda_1 = \frac{W_4}{\tau W_5},\]
where $W_k$ is the $k$th element of $\bW$ for $k=1,\dots,5$.
Therefore, by the delta method, the asymptotic variance of $\log\wR = \log\wlambda_1 - \log\wlambda_0$ can be written as $\bd^{\rm T} var(\bW)\bd$, where 
\begin{align*}
\bd =& \left(-\frac{1}{E\left(W_1\right)}, -\frac{1}{E\left(N-W_2\right)}, \frac{1}{E\left(W_3\right)}, \frac{1}{E\left(W_4\right)}, -\frac{1}{E\left(W_5\right)}\right)^{\rm T}\\
=&\left(-\frac{1}{Np(P_R-\beta_T)}, -\frac{1}{N(1-p)},\frac{1}{\Omega_T-\beta_TT},\frac{1}{N(1-p)r\tau\lambda_1}, -\frac{1}{N(1-p)r}\right)^{\rm T}.
\end{align*}

Note that $N_+\sim Bin(N,p)$, $N_- = N-N_+$,and $N_{-,enroll} \sim Bin (N_-, r)$.
The number of test-recent subjects $N_R$ can be viewed as from $Bin(N_+,P_R)$, where 
\[P_R = \beta_T + \frac{\lambda_0(1-p)}{p}(\Omega_T - \beta_TT).\]
The number of incidence cases $N_{event}$ is from $Poisson (N_{-,enroll}\tau\lambda_1)$.
Then, calculation yields 
\begin{align*}
var(W_1) =&Np\left\{P_R(1-P_R) + (1-p)(P_R - \beta_T)^2 +\sigma^2_{\wbeta_T}(1-p + Np)\right\}\\
var(W_2) =&Np(1-p) \\
var(W_3) =& \sigma_{\wOmega_T}^2 + \sigma_{\wbeta_T}^2T^2\\
var(W_4) =& N(1-p)r\lambda_1\tau\{1+\lambda_1\tau pr + \lambda_1\tau(1-r)\}\\
var(W_5) =& N(1-p)r(1-r+pr)\\
cov(W_1,W_2) =&Np(1-p)(P_R-\beta_T)\\
cov(W_1,W_3) =&Np\sigma_{\wbeta_T}^2T\\
cov(W_1,W_4) =&-Np(1-p)(P_R-\beta_T)r\lambda_1\tau\\
cov(W_1,W_5) =&-Np(1-p)(P_R-\beta_T)r\\
cov(W_2,W_3) =&0\\
cov(W_2,W_4) =&-Np(1-p)r\lambda_1\tau\\
cov(W_2,W_5) =&-Np(1-p)r\\
cov(W_3,W_4) =&0\\
cov(W_3,W_5) =&0\\
cov(W_4,W_5) =&N(1-p)r(1-r+pr)\lambda_1\tau.
\end{align*}

Then, the asymptotic variance of $\log\wR$ is given by $V_0 + V_1 + CV$, where
\begin{align*}
V_0 =&d_1^2var(W_1) + d_3^2var(W_2) + d_4^2var(W_3) + 2d_1d_3cov(W_1,W_2)+ 2d_1d_4cov(W_1,W_3)
\end{align*}
 is the asymptotic variance of $\log\wlambda_0$, 
\[V_1 = d_5^2var(W_4)+d_6^2var(W_5) + 2d_5 d_6 cov(W_4,W_5)\]
is the asymptotic variance of $\log\wlambda_1$, and
\begin{align*}
CV =& 2d_1d_5 cov(W_1,W_4) + 2 d_1d_6 cov(W_1,W_5) + 2d_3d_5 cov(W_2,W_4) + 2d_3d_6 cov(W_2,W_5)
\end{align*}
is the asymptotic covariance of $\log\wlambda_0$ and $\log\wlambda_1$.
Note that 
\begin{align*}
V_0 =&\frac{P_R(1-P_R) +(1-p)(P_R - \beta_T)^2 +\sigma^2_{\wbeta_T}(1-p + Np) }{Np(P_R-\beta_T)^2} + \frac{p}{N(1-p)}\\
&\qquad + \frac{\sigma_{\wOmega_T}^2 + \sigma_{\wbeta_T}^2T^2}{\left(\Omega_T-\beta_TT\right)^2}  + \frac{2}{N}-\frac{2\sigma_{\wbeta_T}^2T}{(P_R-\beta_T)(\Omega_T-\beta_TT)}\\
 =&\frac{1}{Np}\left\{\frac{P_R\left(1-P_R\right)}{\left(P_R-\beta_T\right)^2} +\frac{1}{(1-p)}+\frac{(1-p)\sigma_{\wbeta_T}^2 }{\left(P_R-\beta_T\right)^2}\right\} + \frac{\sigma_{\wOmega_T}^2}{\left(\Omega_T - \beta_TT\right)^2}\\
 &\qquad +\sigma_{\wbeta_T}^2\left\{\frac{\Omega_T - P_RT}{\left(P_R-\beta_T\right)\left(\Omega_T - \beta_TT\right)}\right\}^2,\\
V_1 =& \frac{1+\lambda_1\tau pr + \lambda_1\tau(1-r)}{N(1-p)r\lambda_1\tau} + \frac{1-r+pr}{N(1-p)r}-\frac{2(1-r+pr)}{N(1-p)r} 
 =\frac{1}{N(1-p)r\lambda_1\tau},
 \end{align*}
and 
\[CV = \frac{2}{N} - \frac{2}{N} + \frac{2p}{N(1-p)} - \frac{2p}{N(1-p)} = 0.\]
That is, $\log\lambda_0$ and $\log\lambda_1$ have asymptotic covariance zero and the asymptotic variance of $\wR$ is given in $V_0 + V_1$.
Particularly, the variance of $\log\wlambda_0$ can be estimated by
\begin{align*}
\wV_0 =&\frac{N_R\left(N_+-N_R\right)}{N_+\left(N_R-N_+\wbeta_T\right)^2} +\frac{N}{N_+N_-} + \wsigma_{\wbeta_T}^2\frac{N_+(N-N_+) }{N\left(N_R-N_+\wbeta_T\right)^2} \\
&\qquad + \frac{\wsigma_{\wOmega_T}^2}{\left(\wOmega_T - \wbeta_TT\right)^2}+\wsigma_{\wbeta_T}^2\left\{\frac{N_+\wOmega_T - N_RT}{\left(N_R-N_+\wbeta_T\right)\left(\wOmega_T - \wbeta_TT\right)}\right\}^2,
\end{align*}
 and the variance of $\log\wlambda_1$ can be estimated by
\[\wV_1 =\frac{1}{N_{event}}.\]
In a special case when $\beta_T = 0$ and $\sigma_{\wbeta_T}^2 = 0$, i.e, the false recent probability for the recency test is zero, the variance estimator of $\log\wR$ is given by 
\[\frac{1}{N_R} +\frac{1}{N_-} + \frac{1}{N_{event}} + + \frac{\wsigma_{\wOmega_T}^2}{\wOmega_T^2}.\]
That is, the variance of the estimated incidence ratio is driven by the numbers of observed events and the variability of MDRI of the recency test.
\
 
\section{Derivation of Asymptotic Distribution of $Z$ under Alternatives}\label{append:deriv_Zstat}
%From numerical studies, we see the variance of $Z$ departs from 1 under alternative hypothesis, even with large sample sizes.
In this section, we calculate the asymptotic distribution of $Z$ under alternative hypothesis $R = R_1$.
Particularly, we consider the derivation under a simplified case with $\sigma_{\wOmega_T}^2 = \sigma_{\wbeta_T}^2 = 0$.
Without considering variability associated with the recency assay properties, we will show that the asymptotic variance of $Z$ is a constant (with respect to $N$) that departs from 1 under alternative hypothesis.

Note that $\bW = (N_R - N_+\wbeta_T,N_+, \wOmega_T -\wbeta_TT, N_{event},N_{-,enorll})^{\rm T}$.
In the special case with $\sigma_{\wOmega_T}^2 = \sigma_{\wbeta_T}^2 = 0$, there is no variability associated with $\wbeta_T$ and $\wOmega_T$, such that $\wbeta_T = \beta_T$ and $\wOmega_T = \Omega_T$.
Write $W_6 = N_R$ and $\bW^* = (W_1, W_2, W_4,W_5,W_6)^{\rm T}$.
Then, the test statistic is given by 
\[Z = \frac{A}{\sqrt{B}},\]
where
\begin{align*}
A=&\log\wlambda_1 - \log \wlambda_0 - \log R_0\\
=& -\log(W_1) +\log (N-W_2)+\log (\Omega_T -\beta_TT) +\log W_4 -\log W_5 -\log\tau -\log R_0,\\
B=&\frac{N_R\left(N_+-N_R\right)}{N_+\left(N_R-N_+\beta_T\right)^2} +\frac{N}{N_+N_-} +\frac{1}{N_{event}}\\
=&\frac{W_6\left(W_2-W_6\right)}{W_2W_1^2} +\frac{1}{W_2} +\frac{1}{N-W_2} +\frac{1}{W_4}. 
\end{align*}
We would like to apply the delta method with respect to $\bW^*$ to calculate the distribution of $Z$.

Replacing $W_j$ $(j=1,2,4,5,6)$ by their expectations in the definitions of $A$ and $B$, we denote
\begin{align*}
\tA =&-\log E\left(W_1\right) +\log \{N-E\left(W_2\right)\} +\log (\Omega_T -\beta_TT) +\log E\left(W_4\right)\\
&\qquad -\log E\left(W_5\right) -\log\tau -\log R_0\\
=&\log \lambda_1 -\log\lambda_0- \log R_0\\
\tB =& \frac{E\left(W_6\right)\left\{E\left(W_2\right)-E\left(W_6\right\}\right)}{E\left(W_2\right)E\left(W_1\right)^2} +\frac{1}{E\left(W_2\right)} +\frac{1}{N-E\left(W_2\right)} +\frac{1}{E\left(W_4\right)}\\
=&\frac{1}{N}\left\{\frac{P_R(1-P_R)}{p(P_R - \beta_T)^2} +\frac{1}{p(1-p)} +\frac{1}{(1-p)r\lambda_1\tau}\right\}.
\end{align*}
We apply the delta method to find the asymptotic mean of $Z$ is given by $\tA / \sqrt \tB$,
and the asymptotic variance of $Z$ is given by $\bd_Z^{\rm T} var(\bW^*)\bd_Z$,
where 
\begin{align*}
\bd_Z =& \frac{1}{\sqrt\tB} \left(-\frac{1}{E\left(W_1\right)},-\frac{1}{E\left(N-W_2\right)},\frac{1}{E\left(W_4\right)}, -\frac{1}{E\left(W_5\right)},0\right)^{\rm T}\\
& - \frac{\tA}{2\tB^{3/2}} \left(-\frac{2E\left(W_6\right)\left\{E\left(W_2\right)-E\left(W_6\right)\right\}}{E\left(W_2\right)E\left(W_1\right)^3}
,\frac{E\left(W_6\right)^2}{E\left(W_2\right)^2E\left(W_1\right)^2} -\frac{1}{E\left(W_2\right)^2}-\frac{1}{E\left(N-W_2\right)^2},\right.\\
&\qquad\left. -\frac{1}{E\left(W_4\right)^2},0, \frac{E\left(W_2\right)-2E\left(W_6\right)}{E\left(W_2\right)E\left(W_1\right)^2}\right)^{\rm T}\\
=& \frac{1}{N\sqrt\tB}\bd_{Z1}  - \frac{\tA}{2N^2\tB^{3/2}}\bd_{Z2}.
\end{align*}
where 
\[\bd_{Z1} =\left(-\frac{1}{p(P_R-\beta_T)}, -\frac{1}{1-p},\frac{1}{(1-p)r\tau\lambda_1}, -\frac{1}{(1-p)r},0\right)^{\rm T}\]
and
\[\bd_{Z2 } =\left(-\frac{2P_R(1-P_R)}{p^2(P_R-\beta_T)^3},\frac{P_R^2}{p^2(P_R-\beta_T)^2} -\frac{p^2+(1-p)^2}{p^2(1-p)^2}, -\frac{1}{(1-p)^2r^2\lambda_1^2\tau^2},0,\frac{1-2P_R}{p^2(P_R-\beta_T)^2}\right)^{\rm T}.\]
Since $\tB$ is the variance of $A$, we have $ \bd_{Z1}^{\rm T} var(\bW^*)\bd_{Z1}/N^2=\tB$ and 
\begin{align*}
var(Z) =& \frac{\bd_{Z1}^{\rm T} var(\bW^*)\bd_{Z1}}{N^2\tB} - \frac{\tA\bd_{Z1}^{\rm T} var(\bW^*)\bd_{Z2}}{N^3\tB^2}+\frac{\tA^2\bd_{Z2}^{\rm T} var(\bW^*)\bd_{Z2}}{4N^4\tB^3}\\
=& 1 + \frac{\tA}{4(N\tB)^3}\left\{\tA \bd_{Z2}-4(N\tB) \bd_{Z1}\right\}^{\rm T}\frac{var(\bW^*)}{N}\bd_{Z2}.
\end{align*}

Note that $\tA$ is a constant related to the relationship of $\lambda_0$ and $\lambda_1$.
Particularly, if $\lambda_1/\lambda_0 = R_0$, i.e., the true relationship of $\lambda_0$ and $\lambda_1$ follows from the null hypothesis, then $\tA =0$, $E(Z) = 0$, and 
$var(Z) = 1$.
When $\lambda_1/\lambda_0 = R_1$, i.e., the true relationship of $\lambda_0$ and $\lambda_1$ follows from the alternative hypothesis, $\tA =\log R_1 - \log R_0$, and 
\[E(Z) = \frac{\log R_1 - \log R_0}{\sqrt{V_0(\lambda_0) + V_1(R_1\lambda_0)}}.\]
Note that $var(\bW^*)$ is proportional to $N$ and $\tB$ is proportional to $1/N$.
Then, the second term of the last expression is a constant with respect to $N$.
When this constant is non-zero, the asymptotic variance of $Z$ departs from 1.
Particularly, the asymptotic variance of $Z$ under alternative hypothesis $R=R_1$ is given by the formula
\begin{align*}
V_{R_1} = \bd_{R_1}^{\rm T}V_{\bW}\bd_{R_1},
\end{align*}
where 
\begin{align*}
\bd_{R_1} =& \frac{1}{\sqrt\tB_{R_1}}\bd_{Z1} - \frac{\log R_1 -\log R_0}{2\tB_{R_1}^{3/2}}\bd_{Z2},\\
\tB_{R_1} =& \frac{P_R(1-P_R)}{p(P_R - \beta_T)^2} +\frac{1}{p(1-p)} +\frac{1}{(1-p)r\lambda_0R_1\tau},
\end{align*}
and $V_\bW$ is the covariance matrix of $\bW^*$ divided by $N$ (which does not depend on $N$).
Note that to calculate $V_\bW$, we make use of the covariance matrix of $\bW$ calculated in Appendix \ref{append:deriv_var} and
\begin{align*}
var(W_6) =&NpP_R(1-pP_R),\\
cov(W_1,W_6) =&NpP_R(1-P_R) + Np(1-p)(P_R-\beta_T)P_R,\\
cov(W_2,W_6) =&Np(1-p)P_R,\\
cov(W_4,W_6) =&-Np(1-p)P_Rr\lambda_1\tau,\\
cov(W_5,W_6) =&-Np(1-p)P_Rr.
\end{align*}

$V_{R_1}$ is the calculated asymptotic variance of $Z$ under alternative hypothesis $R=R_1$, in the special case with $\sigma_{\wOmega_T}^2 = \sigma_{\wbeta_T}^2 = 0$.
When the variabilities of $\wbeta_T$ and $\wOmega_T$ cannot be ignored, $V_{R_1}$ serves as an approximation of the asymptotic variance of $Z$.

\end{document}